\documentstyle[preprint,aps]{revtex}
\draft
\begin{document}
\title{Peierls instability and optical response in the 
one--dimensional half--filled Holstein model of spinless fermions}
\author{A. Wei{\ss}e and H. Fehske} 
\address{Physikalisches Institut, Universit\"at Bayreuth,
D--95440 Bayreuth, Germany}
\date{Bayreuth, 22 June 1998}
\maketitle
\input{epsf}
\def\gsim{\hbox{$\lower1pt\hbox{$>$}\above-1pt\raise1pt\hbox{$\sim$}$}}
\def\lsim{\hbox{$\lower1pt\hbox{$<$}\above-1pt\raise1pt\hbox{$\sim$}$}}
\def\cH{{\cal{H}}}
\def\ep{\varepsilon_p}
\def\om{\omega_0}
\def\cS{{\cal{S}}}
\def\mD{{\mit{\Delta}}}
\def\mP{{\mit{\Psi}}}
\begin{abstract}
The effects of quantum lattice fluctuations on the Peierls transition 
are studied within the one--dimensional Holstein molecular crystal model 
by means of exact diagonalization methods.
Applying a very efficient variational Lanczos technique, 
the ground--state phase diagram is obtained in excellent 
agreement with predictions of recent density matrix renormalization 
group calculations.  The transition 
to the charge--density--wave regime is signaled by a strong increase
in the charge structure factor. In the metallic regime, the non--universal 
Luttinger liquid parameters (charge velocity and coupling constant)  
are deduced from a finite--size scaling analysis. 
The variational results are supported 
by a complete numerical solution of the quantum phonon Holstein model 
on small clusters, which is based on a well--controlled phonon Hilbert 
space truncation procedure. The metallic and charge--density--wave phases
are characterized by significant differences in the calculated
optical absorption spectra.  
\end{abstract}
\thispagestyle{empty}
\newpage
\section{Introduction}
Many quasi one--dimensional (1D) materials, 
such as the organic conjugated polymers
[e.g., (CH)$_x$] and charge transfer salts [e.g., TTF(TCNQ)] or 
the inorganic blue bronzes [e.g., $\rm K_{0.3}MoO_3$] and 
MX--chains~\cite{exsum}, undergo a Peierls instability  
in the half--filled band case at temperatures between 50~K and 250~K, 
driven by the electron--phonon (EP) interaction~\cite{Pe55}. 
Most theoretical studies of these systems
concentrate on the 1D SSH~\cite{SSH79} and Holstein~\cite{Ho59a} models, 
where the phonons interact with the electrons by modifying 
the electron hopping matrix element and on--site potential, 
respectively. Frequently the lattice degrees of freedom 
were treated classically. However, it has been argued that for most quasi--1D 
systems the lattice zero--point motion is comparable 
to the Peierls lattice distortion, which makes 
the rigid lattice approximation questionable~\cite{MW92}. 
Although the problem, whether the dimerized ground state
survives the quantum phonon fluctuations, has been addressed
by several numerical~\cite{FH83,HF83,MHM96} and 
analytical~\cite{SZH86,PP88,WHS95} approaches,
it has to date resisted a complete theoretical solution. 
Moreover, lattice dynamical (quantum phonon) effects should be included 
in any theoretical analysis of the extraordinary transport and optical 
phenomena observed in Peierls--distorted 
systems~\cite{HS85,Deea95}. The discussion of 
optical properties, e.g., of the optical absorption, 
poses, however, an extremely complicated many--body problem 
that cannot be solved analytically without 
further approximations~\cite{ZZ96,Geea97a}.    
Perhaps, at present, the most reliable results for models of electrons 
strongly interacting with quantum phonons come from finite--cluster
calculations supplemented by a careful finite--size analysis.

Encouraged by this situation, in this work we present a purely 
numerical (exact diagonalization) study of the 1D spinless fermion 
Holstein model~\cite{spfemo}, 
with a view to understanding the effect of quantum lattice 
fluctuations on the Peierls dimerization and optical absorption
spectra in both the metallic [Luttinger liquid (LL)] and 
insulating [charge--density--wave (CDW)] phases. 
In a particle--hole symmetric notation the Holstein Hamiltonian reads 
\begin{equation}
\cH=-t\sum_i ( c_i^{\dagger} c_{i+1}^{} + c_{i+1}^{\dagger} c_{i}^{} )
- \sqrt{\ep\hbar\omega_0}\sum_i ( b_i^{\dagger}  + b_i^{})\,  (n_i^{} 
-\mbox{\small $\frac{1}{2}$})+
\hbar\om\sum_i  ( b_i^{\dagger} b_i^{}+ \mbox{\small $\frac{1}{2}$})\,, 
\label{homo}
\end{equation} 
where $c_i^{[\dagger]}$ ($b_i^{[\dagger]}$) are
the electron (phonon) annihilation [creation] operators, 
and $n_i^{}=c_i^{\dagger} c_i^{}$.  
In Eq.~(\ref{homo}), the free--electron transfer amplitude
$t$ is restricted to nearest--neighbour hopping,  
a dispersionsless Einstein phonon  $\omega(q)=\omega_0$ 
is coupled to the local electron density, 
and the phonons are treated within harmonic approximation. 
In the atomic limit ($t=0$), $\ep$ gives the well--known 
Lang--Firsov polaron binding energy. Rescaling $\cH\to\cH/t$
and measuring all energies in units of $t$, it is convenient
to introduce the adiabaticity parameter $\alpha=\hbar\omega_0/t$
and two dimensionless EP coupling constants, $\lambda=\ep/2t$
and $g=\sqrt{\ep/\hbar\omega_0}$, in order to characterize the weak 
$(\lambda\ll 1)$ and strong coupling ($\lambda\gg 1$ {\it and} $g\gg1$) 
situations in the adiabatic ($\alpha \ll 1$) and anti--adiabatic 
($\alpha \gg 1$) regimes. For the single--carrier case, the Holstein 
Hamiltonian was extensively studied in the context of the polaron problem. 
A practically complete numerical solution, including ground state 
and spectral properties, is now available 
from ED (exact diagonalization) and DMRG 
(density matrix renormalization group) 
calculations~\cite{EDHomo,FLW97,WF97}.
At half--filling, a number of different analytical and numerical methods, 
including strong coupling expansions~\cite{HF83}, 
variational approaches~\cite{ZFA89} and renormalization group 
arguments~\cite{CB84}, as well as
world--line quantum Monte Carlo (WL~QMC)~\cite{HF83}, 
Green's function Monte Carlo (GF~MC)~\cite{MHM96} and DMRG 
techniques~\cite{BMH98}, have been used to determine
the phase boundary between metallic and insulating behaviour.

In this paper, we follow our strategy pursued in recent work
on the self--trapping problem in the single--electron Holstein 
model~\cite{WF98}, and apply two different numerical methods:
(i) a complete exact diagonalization (ED) of the Holstein model preserving
the full dynamics and quantum nature of phonons and (ii) a 
variational Lanczos scheme based on the inhomogeneous modified 
variational Lang--Firsov transformation (IMVLF)~\cite{FRWM95}.
It is natural for the first method to be limited to rather small
clusters (with $N$ sites) and moderate EP coupling 
strengths ($\lambda,\;g)$, mainly due to the necessity of 
a phonon Hilbert space truncation (retaining at most $M$ phonons; 
for details of the numerical method see Ref.~\cite{BWF98}). 
By combining ED with Chebyshev recursion and maximum entropy 
methods~\cite{Siea96,BWF98}, we are able to discuss the 
dynamical properties of the spinless fermion Holstein model, 
such as the optical conductivity (see Sec.~IV).
On the other hand, using the second method, we can study the 
ground--state properties of fairly large systems  (Sec.~II),
which enables us to carry out a finite--size scaling (cf. Sec.~III).
Within the IMVLF--Lanczos approach, we treat the phonon subsystem 
by performing first of all a canonical transformation, 
$\tilde{\cH} = {\cal{U}}^\dagger\, \cH \, {\cal{U}}$, 
${\cal{U}}= \mbox{e}^{-\cS_1(\mD_i)}\,  \mbox{e}^{-\cS_2(\gamma)}\,  
\mbox{e}^{-\cS_3(\tau)}$, where     
$\cS_{1}(\mD_i)= - \frac{1}{2 g\alpha} \sum_i  \mD_i
( b_i^{\dagger} -  b_i^{} )$, 
$\cS_{2}(\bar{\gamma},\gamma)=- g \sum_{i} 
(b_i^{\dagger} -  b_i^{})\,(\bar{\gamma}+\gamma n_i^{})$, and  
$\cS_{3}(\tau)={1\over 2} \ln \tau \sum_i (b_i^{\dagger} b_i^{\dagger} 
- b_i^{} b_i^{} )$
are designed to describe static displacement field $(\mD_i)$, non--adiabatic
polaron $(\gamma,\bar{\gamma})$, and squeezing  $(\tau)$ effects, respectively.
Next, we approximate the eigenstates $|\tilde{\mP} \rangle$ of 
the transformed Hamiltonian by the variational product states 
$|\tilde{\mP}_V \rangle = |\tilde{\mP}_{ph} \rangle 
\otimes|\tilde{\mP}_{el} \rangle$ and average $\tilde{\cH}$ 
over the phonon vacuum, $\bar{\cH}\equiv \langle\tilde{\mP}_{ph}^0 
|\tilde{\cH} |\tilde{\mit \Psi}_{ph}^0 \rangle$, which leads to  
an effective electronic Hamiltonian 
\begin{eqnarray}
\label{heff}
 \bar{\cH}  &=& \, g^2 \alpha(\bar{\gamma^2}+\bar{\gamma}) N \,+\, 
 g^2 \alpha [\gamma^2-\gamma + 2\bar\gamma(\gamma-1)] \sum_i n_i^{} 
\,-\,\mbox{e}^{-g^2\gamma^2\tau^2}
\sum_i (c_i^{\dagger} c_{i+1}^{} + c_{i+1}^{\dagger} c_i^{})    
\nonumber\\[0.2cm]
&&-\, (1-\gamma)  \sum_{i} \mD_i^{}\, (n_i^{} - \mbox{\small $\frac{1}{2}$} ) 
\,+ \eta \sum_i \mD_i 
\,+\,\frac{\alpha N}{4} (\tau^2+\tau^{-2})\,
+\,\frac{1}{4g^2\alpha} \sum_i \mD_i^2\,. 
\end{eqnarray}
Here $\eta$ is a Lagrange multiplier ensuring the constraint 
$ \sum_{i} \mD_i^{}=0$. 
Employing the Hellmann--Feynman theorem, the $N+3$ variational 
parameters are obtained by iteratively solving the extremal 
equations for the corresponding energy functional 
$\bar{E}_0(\{\mD_i\},\bar{\gamma},\gamma,\tau^2)$ in 
combination with the Lanczos recursion algorithm. 

\section{Phase diagram}
Previous results for the ground--state
phase diagram of the Holstein model at half--filling 
obtained by WL QMC~\cite{HF83} and GF MC~\cite{MHM96} simulations
showed significant discrepancies in the region of small $\alpha$ 
($0 < \alpha \lesssim 1$).
Only very recently Bursill et al.~\cite{BMH98} provided more reliable 
information from level crossings in their DMRG data.
Applying, in a first step, our variational Lanczos scheme, we consider 
the effective model~(\ref{heff})
on chains of even length with up to 16 sites and periodic (antiperiodic) 
boundary conditions if there is an odd (even) number of fermions in the
system. To elude the problem of trapping in metastable
minima of the energy functional $\bar{E}_0$, 
we start the variational Lanczos iteration  
with different initial configurations $\{\Delta_i,\,\bar{\gamma},\,
\gamma,\,\tau^2\}$, close to the metallic or the dimerized phase.
As one might expect, in the dimerized phase the iteration always converges
to a ground--state with staggered dimerization $\Delta_i = \Delta (-1)^i$, 
while in the metallic phase $\Delta_i = 0$.
Thus, at half--filling, our inhomogeneous variational wave function
becomes exactly the staggered (SMVLF) one used in Ref.~\cite{ZFA89}, and 
the effective Hamiltonian $\bar{\mathcal H}(\Delta,\gamma,\tau^2)$ can be 
solved easily, also for the infinite system. However, the phase
diagram given in~\cite{ZFA89} for the infinite system is only tentative,
and for small $\alpha$ the determination of $g_c$ is not clear.
Moreover the infinite system is never really gapless within the
variational approach, because $\Delta$ remains nonzero, although it becomes
very small for weak EP coupling.
This situation changes as soon as the system is finite. The dimerization
$\Delta$ then switches from zero to a finite value at a critical coupling
$g_c(\alpha,N)$, where $g_c$ is nearly independend from the system--size~$N$
for large $\alpha$ ($\alpha \gtrsim 1$), but decreases with $N$ for small 
$\alpha$. 

Proceeding this way we get the IMVLF transition lines $g_c(\alpha,N)$
depicted in Fig.~1. Most notably we found that the IMVLF 
phase boundary, separating metallic (LL) and insulating (CDW) phases,
in the whole parameter regime, agrees surprisingly well with 
the very recent DMRG results~\cite{BMH98} (open squares). 
Also in the phase diagram are the transition points obtained by  
WL~QMC~\cite{HF83} and GF~MC~\cite{MHM96}.    

In a second step, performing exact diagonalizations of the full
Hamiltonian for systems with~6 and~10 sites and up to 30 phonons,
we want to demonstrate that the phase boundary determined 
from the effective model~(\ref{heff}) is consistent with what is obtained
for the quantum phonon Holstein model~(\ref{homo}). To this end, 
we have calculated the static charge structure factor, 
\begin{equation}
\label{chi}
\chi(\pi)={1 \over N} \sum_{i,j}\,\mbox{e}^{i\pi(R_i-R_j)}\,\langle 
n_in_j\rangle\,,
\end{equation}
shown in Fig.~2 as a function of $g$ 
at low~(a), intermediate~(b), and high~(c) phonon frequencies for both, the
variational and the exact solution.
Increasing the EP coupling at fixed phonon frequency, the smooth variation
of $\chi (\pi)$ in the metallic phase is followed by a strong enhancement
at about $g_c^{\rm(IMVLF)}$, unambiguously indicating the formation of a CDW.
The discontinuous jump--like behaviour of $\chi^{\rm(IMVLF)}(\pi)$ at $g_c$
is an apparent shortcoming of the variational approach. This resolves also
the open question in~\cite{ZFA89}, whether the two--minimum structure of
the variational solution at large $\alpha$ is an artifact.
For $\alpha\to 0$ and $N\to\infty$, where the IMVLF approach becomes exact,
we found a continuous crossover in  $\chi^{\rm(IMVLF)}(\pi)$ as well. 

Obviously,
in the CDW--like phase, a larger number of phonons ($M$) 
is required to achieve a satisfactory convergence of the ED data 
(see Fig.~2 (b)).
Furthermore, it is interesting to compare the behaviour of the kinetic
energy $\langle E_{kin}\rangle$, given by the average of the first term of~(\ref{homo}),
in the adiabatic, non--adiabatic and anti--adiabatic regimes (see insets).    
For the adiabatic case, the kinetic energy is only weakly reduced from its
noninteracting value  ($\langle E_{kin}\rangle=-4$) in the metallic phase. 
By contrast, in the anti-adiabatic regime, we observe 
a strong reduction of $\langle E_{kin}\rangle$, which can be 
attributed to the formation of a strongly correlated polaronic metal 
{\it below} the CDW transition point.

Coming back to the phase diagram, 
in the {\it adiabatic regime}, our results seem to confirm 
that there is no long--range order for sufficiently small EP coupling, 
which is consistent with the predictions of Refs.~\cite{BGL95,MHM96}. 
At $\alpha=0$, the critical coupling converges
to zero, as expected for the adiabatic Hamiltonian 
($M\to\infty$; $\gamma=0,\, \tau^2=1$).
In the regime $0< \alpha\stackrel{<}{\sim}1$, however, 
the precise determination of $g_c$ is somewhat difficult.  
Maybe the discrepancy between 
the predictions of IMVLF, GF~MC, DMRG on one side and WL MC on the 
other side, of how the critical $\lambda_c$ scales to zero 
with $\alpha\to 0$, results from this ambiguity.

In the strong--coupling {\it non--adiabatic regime} ($g^2,\,\lambda \gg 1$),
the results of the different numerical approaches approximately agree.  
Also the analytical approach, giving the exactly soluble XXZ 
model~\cite{HF83,YY66} within second order perturbation theory 
(with respect to $t$)
\begin{equation}
\label{xxz}  
H^{\rm XXZ} = {N\over 4}(2\alpha-g^2\alpha -V_2) -\mbox{e}^{-g^2} \sum_i\Big(
          (S_i^+ S_{i+1}^- + S_i^- S_{i+1}^+) -
          V_2 \,\mbox{e}^{g^2}  S_i^z S_{i+1}^z \Big)
\end{equation}
with
\begin{equation}
 V_n(\alpha,g^2)= {2 e^{-ng^2} \over\alpha}\sum_{s\ne0} {(ng^2)^s\over s s!}\,,
\end{equation}
works very well. The (Kosterlitz-Thouless) phase transition line is given 
by the condition
$V_2(\alpha,g^2)\mbox{e}^{g^2}/2=1$ (dashed curve in Fig.~1). 
For $\alpha\to \infty$  (anti--adiabatic limit) there is no dimerization 
if $\lambda$ is finite.

To get a feeling about the accuracy of the different analytical and
numerical techniques, we have compared in Fig.~3 
the ground--state energies at high phonon frequencies, 
where the small polaron approximation is justified. 
For the adiabaticity ratio $\alpha=10$, 
used in Fig.~3, the transition to the CDW phase takes 
place at about $g_c\simeq 2$.

Restricting ourselves to the metallic
phase, the ground--state energy of the 
XXZ model is easy to evaluate. It is (per site) 
\begin{equation}
\label{e0xxz}
E_0^{\rm XXZ}={\alpha\over 2} - {g^2\alpha\over 4} 
-\sin[\mu] \,\mbox{e}^{-g^2} 
\int_{-\infty}^{\infty}dx \frac{\mbox{sinh}[(\pi-\mu)x]}{\mbox{sinh}[\pi x]\,
\mbox{sinh}[\mu x]}
\end{equation}
with $\cos [\mu] = V_2\,\mbox{e}^{g^2}/2$. 
Alternatively, calculating the polaron self--energy of the 
Holstein model~(\ref{homo}) within standard  
second--order (Rayleigh--Schr\"odinger)  
strong--coupling perturbation theory (SCPT) and omitting the residual 
polaron interaction, the small polaron band dispersion becomes~\cite{FLW97} 
\begin{equation}
\label{scpt}
  E_K^{\rm SCPT}= - {g^2\alpha\over 2} + V_2
         - 2 \mbox{e}^{-g^2} \cos K - 
         V_1 \,\mbox{e}^{-g^2}\cos 2K\,.
\end{equation}
Then, for the half--filled band case, the ground--state energy (per site) 
takes the form 
\begin{equation}
\label{e0scpt}
E_0^{\rm SCPT} =  E_0^{(1)}  + {V_2\over 2}\,,
\end{equation}
where the (first--order) polaron ground--state energy,  
\begin{equation}
\label{e01}
E_0^{(1)}= {\alpha\over 2}-{g^2\alpha\over 4} - 2 \mbox{e}^{-g^2} /\pi\,,
\end{equation}
can be obtained from~(\ref{heff}) by setting
$\mD_i=0$, $\gamma=1$, $\bar{\gamma}=-1/2$, and $\tau^2=1$.

Both $E_0^{\rm XXZ}$ and $E_0^{\rm SCPT}$ are also depicted in Fig.~3, 
and in order to visualize the higher--order corrections  
we have shifted all energies by the standard small polaron term $E_0^{(1)}$.
First of all we see that the higher--order corrections, originated   
by the residual polaron--phonon interaction, are most important at
intermediate couplings $g\simeq 1$, where the polaron band structure
significantly deviates from a rescaled cosine 
tight--binding band~\cite{FLW97}.  As already mentioned above,
the IMVLF--Lanczos results, extrapolated to $N=\infty$, coincide 
with the variational SMVLF solution~\cite{ZFA89},
but give higher ground--state energies than the XXZ model 
in the intermediate--to--strong  coupling regime. 
This is because the non--adiabatic polaron effects are 
only included to the lowest order of approximation 
(remind that the model (\ref{heff}) was obtained performing 
the average over the zero--phonon state). 
Including second order corrections by SCPT, we found a much better agreement
with the exact data (full circles). Notice the substantial lowering of the 
energy with respect to the first order result at large EP couplings
[$g=2$ implies $\lambda=20$ (!)], which results  
from the momentum independent shift in the kinetic energy~\cite{WF97,FK97}.   
As expected, the finite--size effects due to the lattice discreteness 
are most pronounced in the weak--coupling regime. 

\section{Luttinger liquid parameters}
According to Haldane's Luttinger liquid conjecture~\cite{Ha80}, 1D 
gapless systems of interacting fermions should belong
to the same universality class as the Tomonaga-Luttinger model.
As stated above, the Holstein system is gapless for small 
enough coupling $g$. Thus it is obvious to prove, 
following the lines of approach to the problem by 
McKenzie~et~al.~\cite{MHM96}, whether our IMVLF--Lanczos data 
shows a finite--size scaling like a Luttinger liquid. 

For a LL of spinless fermions, the ground--state 
energy $E_0(N)$ of a finite system of $N$ sites 
scales to leading order as~\cite{Vo95}
\begin{equation}
\label{uro}
  {E_0(N)\over N} = \epsilon_{\infty} - {\pi u_{\rho}\over 6 N^2} \,,
\end{equation}
where $\epsilon_{\infty}$ denotes the ground--state energy per site for
the infinite system and $u_{\rho}$ is the velocity of the charge excitations.
If $E_{\pm 1}(N)$ is the ground--state energy with $\pm 1$ fermions
away from half filling, to leading order the scaling should be
\begin{equation}
\label{kro}
  E_{\pm 1}(N) - E_0(N) = {\pi u_{\rho} \over 2 K_{\rho} N}\,.
\end{equation}
$K_{\rho}$ is  the renormalized effective coupling (stiffness)
constant (for a more detailed discussion see Ref.~\cite{MHM96}).

As becomes evident from Fig.~4, our IMVLF--Lanczos data match both 
scaling relations for all regimes of parameters $\alpha$ and $g$ 
with great accuracy. Let us stress that the IMVLF--energies 
for $E_{\pm 1}(N)$ correspond to a inhomogeneous solution for
the displacement fields $\mD_i$, which deviates from a uniform or 
staggered ordering, and therefore cannot be obtained within the 
simple SMVLF scheme~\cite{ZFA89}.

In the plot of the LL parameters, Fig.~5, increasing error bars at
the end of the curves indicate the phase transition to the CDW state.
Here the scaling relations~(\ref{uro}) and~(\ref{kro}) no longer hold.  
Perhaps surprising, the velocity of charge excitations 
$u_{\rho}$, reflecting the behaviour of the 
kinetic energy (cf.~Fig.~2 insets), agrees fairly well with the results of 
McKenzie~et~al.~\cite{MHM96}, while the coupling constant $K_{\rho}$
is always bigger than one, indicating an attractive interaction.
We believe this is an artifact of our variational treatment. For the 
XXZ model $K_\rho \to 1/2$, i.e. the phase transition is of infinite
order~\cite{Sh90} with a Kosterlitz--Thouless order parameter 
$\sim e^{-1/(g-g_c)}$. Note that
the parameters $u_{\rho}$ and $K_{\rho}$ we show, are of course those
for the effective Hamiltonian~(\ref{heff}), which coincide with the
parameters of the true Holstein model only more or less.

Unfortunately, extracting a similar scaling behaviour 
from our ED data seems to be extremely complicated, mainly  
because {\it two} finite--size dependences, those with respect to 
the system size $N$ and maximum phonon number $M$, are mixed.
Moreover, the memory limitations of the present day parallel computers 
impose severe restrictions on the lattice sizes ($N\leq 10$), 
that can be treated by ED with adequate accuracy.

\section{Optical response}
One of the physical quantities which contains extremely valuable 
information about the low--energy excitations
in polaronic metals and CDW systems is the optical conductivity, 
$\sigma(\omega)$, usually determined
from reflectivity measurements. As will be shown in the following,
the optical absorption spectra of nearly free electrons, small polarons
and CDW insulators differ essentially. 

The real part of $\sigma(\omega)$ contains two contributions, the familiar  
(coherent) Drude part at $\omega=0$ and a so--called ``regular term'',
$\sigma^{reg}(\omega)$, due to finite--frequency dissipative optical
transitions to excited quasiparticle states. In spectral representation 
($T=0$), the regular part takes the form~\cite{BWF98,Da94}
\begin{equation}
\label{sigreg} 
\sigma^{reg}(\omega)=\sum_{m > 0}
\frac{|\langle {\mit \Psi}_0^{} |i \sum_{j}( c_{j}^{\dagger}
 c_{j+1}^{} - c_{j+1}^{\dagger}c_{j}^{}) |  {\mit \Psi}_m^{} 
       \rangle |^2}{E_m-E_0} \;\delta[\omega -(E_m-E_0)]\,,
\end{equation}
where $\sigma^{reg}(\omega)$ is given in units 
of $\pi e^2$ and we have omitted an $1/N$ prefactor. 
In~(\ref{sigreg}), the summation is taken over the complete set 
of eigenstates $|{\mit \Psi}_m^{}\rangle$ with excitation 
energies $\omega =(E_m-E_0)$ in the subspace of $N/2$ (spinless) 
fermions (half--filling). 
For the discussion of the optical properties it is useful 
to consider also the $\omega$--integrated spectral weight function  
\begin{equation}
\cS^{reg}(\omega)=\int_0^{\omega}d\omega^{\prime}\sigma^{reg}
(\omega^{\prime})\,.
\end{equation}
The evaluation of dynamical
correlation functions, such as~(\ref{sigreg}), can be carried out 
by means of very efficient and numerically stable Chebyshev recursion and 
maximum entropy algorithms~\cite{Siea96,BWF98}. Nevertheless, due to 
the huge size of the Hilbert space in this EP model, 
we are currently restricted to a lattice size of 6 
sites ($M=30$; periodic boundary conditions) 
if we want to calculate the conductivity in  a wide range of
EP coupling strengths.

Typical optical absorption spectra for the 1D half--filled Holstein model
of spinless fermions are given by Figs.~6, 7, and~8, in 
the metallic~(a) and CDW~(b) phases, at characteristic phonon 
frequencies corresponding to the adiabatic, intermediate, 
and anti--adiabatic regimes, respectively. 

For {\it low phonon frequencies} ($\alpha\ll 1$) and weak EP couplings 
(see Fig.~6~a), the peak structure may be easily understood in connection with
the non--interacting tight--binding band dispersion $E_K^{(0)}=-2 \cos K$,
where the allowed $K$ values are $K=0$, $\pm \pi/3$, $\pm 2\pi/3$, and $\pi$
for a six site system.  
Obviously, we found the first transitions with non--negligible (electronic) 
spectral weight (cf. $S^{reg}(\omega)$) at frequencies that 
approximately correspond to to the discrete free electron Bloch states 
of our finite system and its vibrational satellites. Accordingly the first 
and second group of excitations originate from transitions where 
the momentum of one electron is changed from  
$\pm \pi/3$ to $\pm 2\pi/3$ and $\pm \pi/3$ to $\pm \pi$, respectively. 
Note that in~(\ref{sigreg}) an optical transition only takes 
place within the $K=0$ sector ($|\mP_0\rangle$ carries $K=0$ for 
the half--filled band case). Thus a phonon with opposite 
momentum must be absorbed in order to ensure momentum
conservation during a single--particle excitation process. Of course, in the 
Holstein model, $K$ is the {\it total} momentum of the coupled EP system,
and, at any $g>0$, there is a finite overlap of the ground state 
with all the excited states belonging to the same $K$ sector. 
The most relevant point is that in the metallic phase the
absorption threshold should tend to zero as the number 
of sites increases, i.e., the low--energy (finite--size) gap vanishes.

The optical absorption spectrum in the strong EP coupling regime
is quite different from that in the LL phase. 
It can be interpreted in terms of strong electron--phonon correlations 
and corroborates the CDW picture. For $g>g_c$ the electronic 
band structure is gapped (at the edge points $K=\pm \pi/2$), 
and we expect that now the low--energy gap feature, observed in Fig.~6~(b),
survives in the thermodynamic limit $N\to\infty$. 
Unfortunately finite--size effects prevent a precise extraction of the CDW gap
from our optical ED data. The broad optical absorption band found
above this gap is produced by a single-particle excitation accompanied
by multi--phonon absorptions and is basically related to 
the lowest unoccupied state of the upper band of the CDW insulator. 
The lineshape reflects the phonon distribution in the ground state 
(see Fig.~9).  
The most striking feature is the strong increase of the spectral weight
contained in the incoherent part of optical conductivity. This becomes evident
by comparing the magnitude of $S^{reg}(\infty)$ in the weak and strong 
coupling situations. Moreover, employing the f--sum rule for 
the optical conductivity~\cite{WF98} and taking into account 
the behaviour of the kinetic energy as function of $g$ (see Fig.~3), 
we found that in the metallic and CDW phases nearly all the spectral weight
is contained in the coherent (Drude) and incoherent (regular) part 
of $\mbox{Re}\ \sigma(\omega)$, respectively. 
That is, in the CDW state the transport is dominated 
by inelastic scattering processes.     

In the {\it non--adiabatic region} $\alpha\simeq 1$, 
where the phonon frequency becomes comparable to the electronic 
bandwidth (level spacing), the situation is not much different (see Fig.~7). 
Again, in the weakly interacting case, the optical absorption
can be understood in terms of electronic transitions within a tight--binding
band and phonon satellites.
Crossing the CDW transition point, a pronounced redistribution 
of spectral weight from the Drude to the regular part 
of $\sigma(\omega)$ is observed. 
As can be seen by comparing Figs.~6--8 and Fig.~9,  the heights 
of the jumps in the $\omega$--integrated conductivity, being directly related
to the probability of the corresponding $m$--phonon absorption processes,
give a measure of the weights of the $m$--phonon states in the ground state.

Finally, we consider the optical response in the {\it anti--adiabatic regime}. 
For weak interaction the picture remains the same as in the above 
adiabatic and intermediate cases (see Fig.~8).  
If we increase
the EP coupling $g$, the electrons will be heavily dressed by the phonons
(which now can follow the electron instantaneously) and the formation
of less mobile small polarons takes place (cf. $\langle E_{kin}(g)\rangle$ 
and $u_{\rho}(g)$
shown in Fig.~3~(c) and Fig.~5~(a), respectively). As a consequence 
the coherent transport becomes strongly suppressed in the LL phase. 
Since the (renormalized) coherent bandwidth of the polaron band 
is rather small, the finite--size gaps in the band structure
are reduced as well, and the CDW gap ($\mD_{CDW}\sim 2\lambda$)   
may be identified with the optical absorption threshold.
Again we found that in the CDW phase multi--phonon absorption
processes dominate the optical response. 

\section{Summary}
In this paper we have studied the Peierls instability
and the optical absorption in the half--filled spinless 
fermion Holstein model by means of finite--lattice diagonalizations.
We have shown that the simple variational (IMVLF) Lanczos approach 
can be successfully used to determine the ground--state properties
of the Holstein model, in particular the phase diagram.
The calculation of the optical properties did require a 
complete diagonalization of the model, preserving the full
dynamics of the phonons.

Our results confirm previous findings that at
sufficiently weak electron--phonon (EP) interaction  
the system resides in a metallic (gapless) 
phase, described by two Luttinger liquid parameters.
The renormalized charge velocity ($u_{\rho}$) and correlation exponent
($K_{\rho}$) were obtained from finite--size scaling relations, 
fulfilled with great accuracy. Increasing the EP coupling, the 
system undergoes a Peierls transition to an insulating (gaped) phase,
reflected in a strong increase of the charge--density correlations
with momentum $\pi$. The crossover between Luttinger liquid and 
charge--density--density (CDW) behaviour is found in good
agreement with exact diagonalization and density matrix renormalization
results for the quantum phonon model. In the non--adiabatic 
strong--coupling limit, where the charge carriers are polaronic, 
the IMVLF--Lanczos phase boundary lies close to  
the analytic findings for the XXZ (small polaron) model.  
The transition to the CDW state is accompanied by significant changes
in the optical response of the system. Most notably seems 
to be the substantial spectral weight transfer 
from the Drude to the regular (incoherent) part of the optical conductivity, 
indicating the increasing importance of inelastic scattering processes 
in the CDW (Peierls distorted) regime.

The numerical results made clear that a dynamical treatment of the 
lattice degrees is necessary in the intermediate frequency
and coupling region because the energy scales are not well separated.
This should be a matter of relative importance modeling 
the inorganic spin Peierls materials~\cite{xyco}, e.g. $\rm CuGeO_3$, 
where the spin exchange interaction and the relevant 
phonon frequencies are of the same order.  
Indeed, applying our numerical techniques to a frustrated Heisenberg 
spin--$1\over 2$ chain with dynamic spin--phonon coupling shows that in
the non--adiabatic regime the spin--Peierls transition takes place at
about $g_c \sim 1$, which  implies for $\rm CuGeO_3$ a intermediate to strong
coupling situation~\cite{WFK98}.

\acknowledgments
The authors would like to thank G. Wellein and B. B\"auml 
for their support in the numerical work.
We are particularly indebted to R. J. Bursill for putting his 
DMRG data prior publication at our disposal, and to E.~Jeckelmann for useful
comments on the optical conductivity data.  
Calculations were performed at the LRZ M\"unchen, HLRZ J\"ulich, and the HLR 
Stuttgart.
\narrowtext\def\baselinestretch{0.95}
\bibliography{ref}
\bibliographystyle{phys}
\figure{FIG. 1. Ground--state phase diagram of the 1D Holstein model 
of spinless fermions at half filling, showing the boundary 
between the Luttinger liquid (LL) and charge--density--wave (CDW) states. 
The IMVLF--Lanczos results are compared with the predictions of 
different analytical and numerical approaches.} 
\figure{FIG. 2. Charge structure factor $\chi(\pi)$ and kinetic energy
$\langle E_{kin}\rangle$ (inset) as a function of the EP 
coupling $g$ in the adiabatic (a), non--adiabatic (b), 
and anti--adiabatic (c) regimes.}
\figure{FIG. 3. Dependence on the EP coupling of 
the ground state energy (shifted by $E_0^{(1)}$).
The predictions of strong coupling expansions are confronted with 
IMVLF--Lanczos and ED results obtained for a six--site lattice at $\alpha=10$.}
\figure{FIG. 4. Finite--size scaling of the charge gap $E_{-1}(N) - E_0(N)$
and the ground state energy $E_0(N)$ (inset) for different values of $g$  
($g^2=0.1$, 0.6, 1.1, 1.6, 2.0 from top to bottom) at $\alpha=0.1$ (a), 1 
(b), and 10 (c).}
\figure{FIG. 5. LL parameters  $u_{\rho}$ [charge velocity (a)] and 
$K_{\rho}$  [correlation exponent (b)] as a function of the 
EP coupling $g$. The dashed curves denote the corresponding results 
for the XXZ model at $\alpha=10$.} 
\figure{FIG. 6. Regular part of the optical conductivity 
$\sigma^{reg}(\omega)$ (dotted line) and integrated spectral weight
$S^{reg}(\omega)$ in the adiabatic weak (a) and strong (b) 
EP coupling regimes ($N=6$, $M=30$.}
\figure{FIG. 7. The same as Fig. 6 but at $\alpha=1$.
The results in the LL and CDW regimes are shown an (a) and (b), respectively.}
\figure{FIG. 8. Optical absorption in the anti--adiabatic regime ($\alpha=10$)
of the 1D half--filled Holstein model.}
\figure{FIG. 9. Phonon--distribution function $|c^m|^2$, as defined in 
equation (A9) of~\cite{BWF98}, shown for the parameters used in FIG.~6--8.}
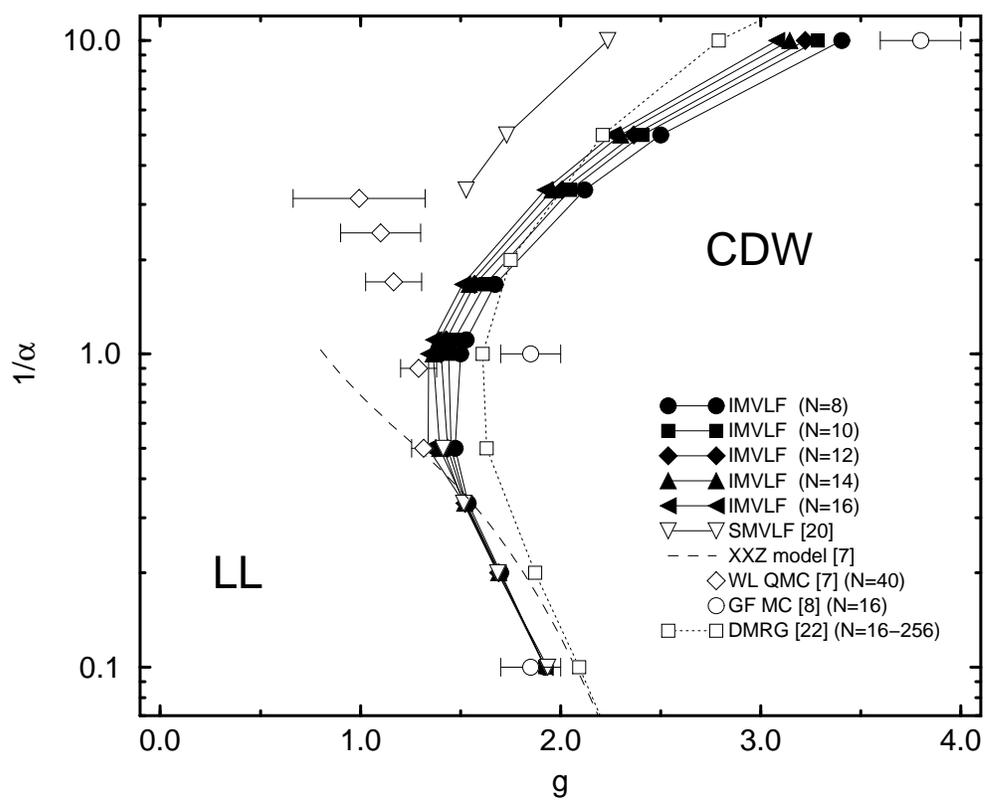
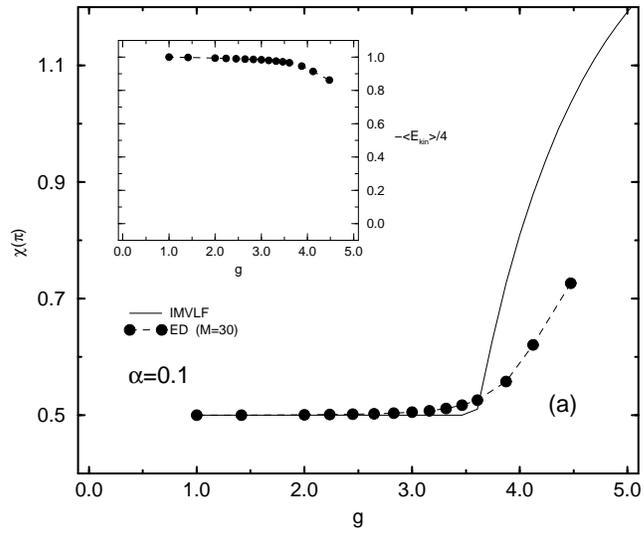
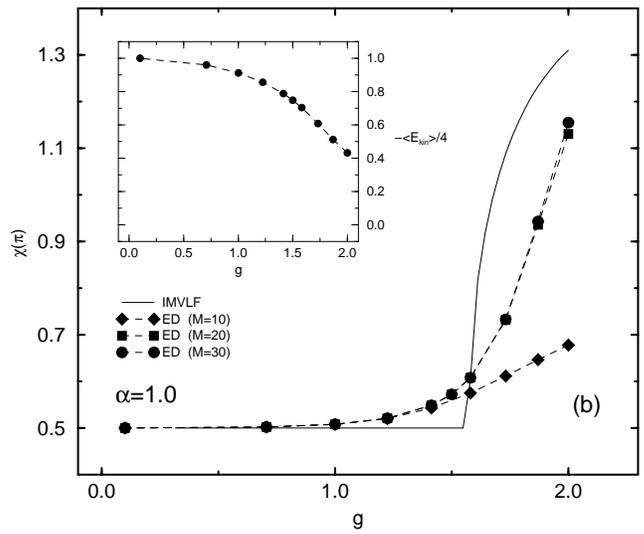
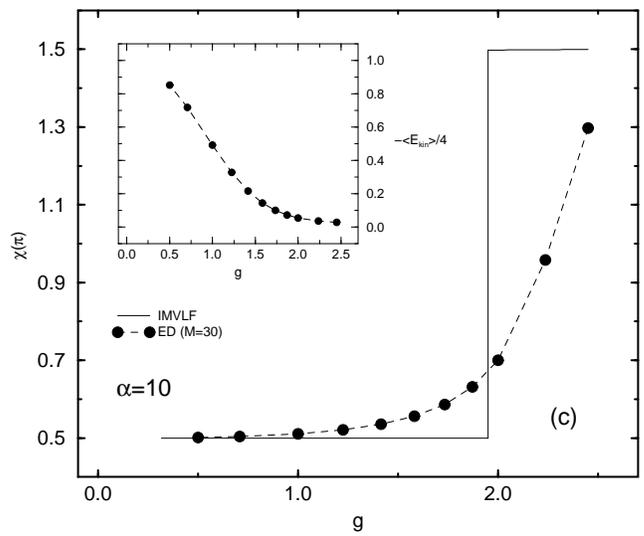
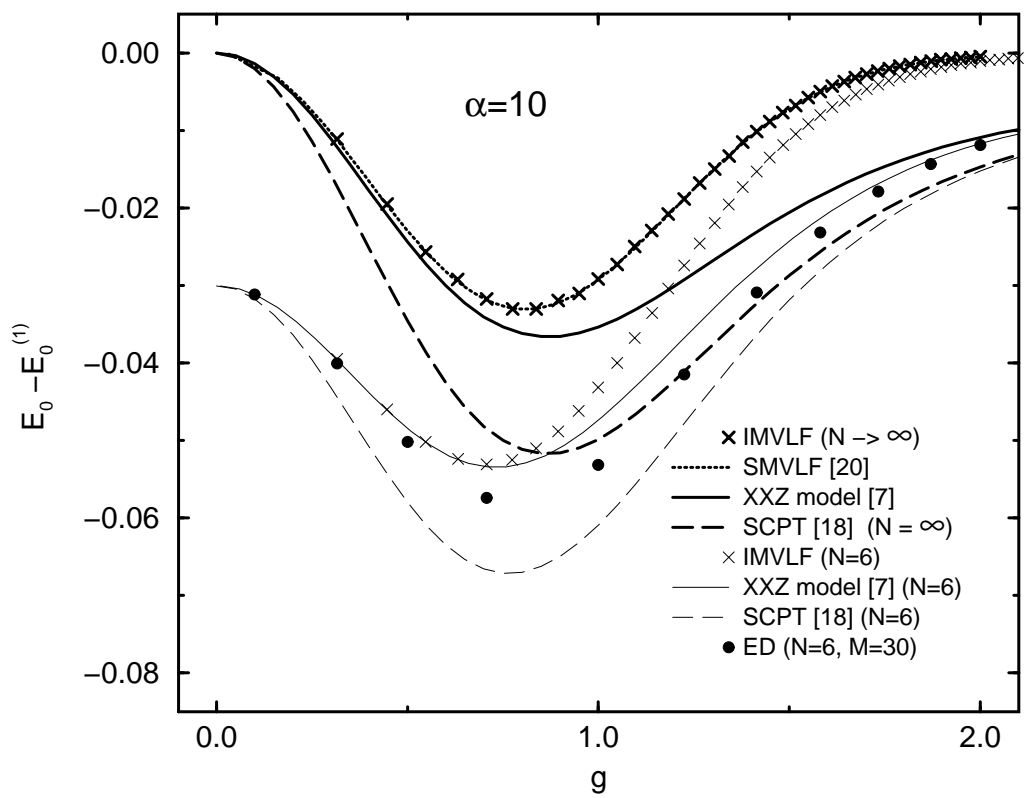
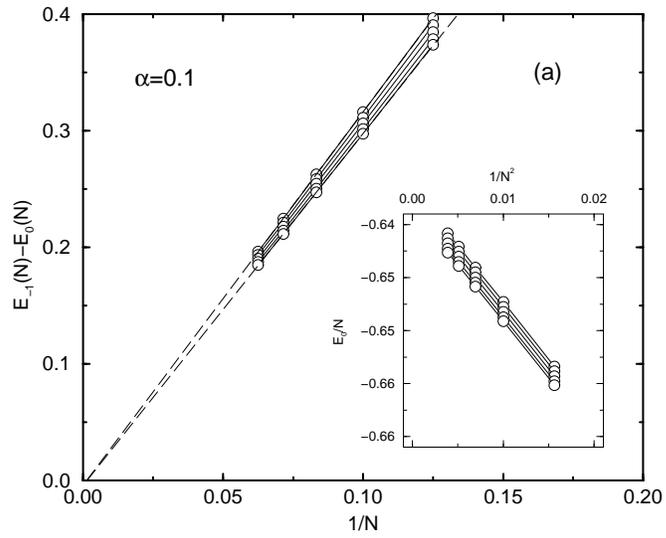
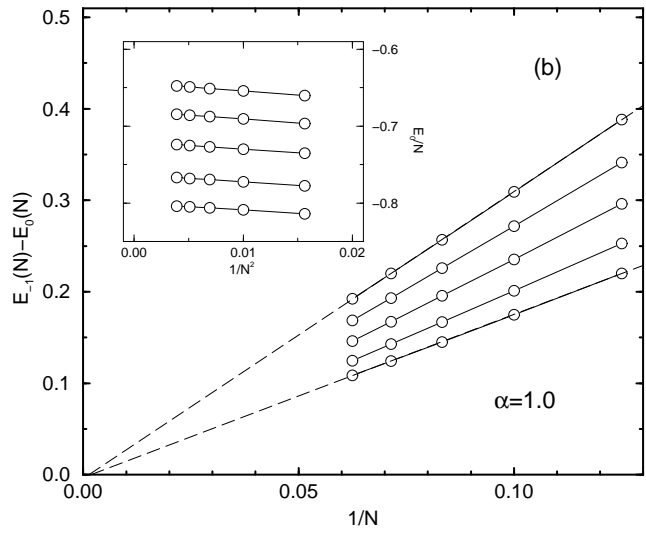
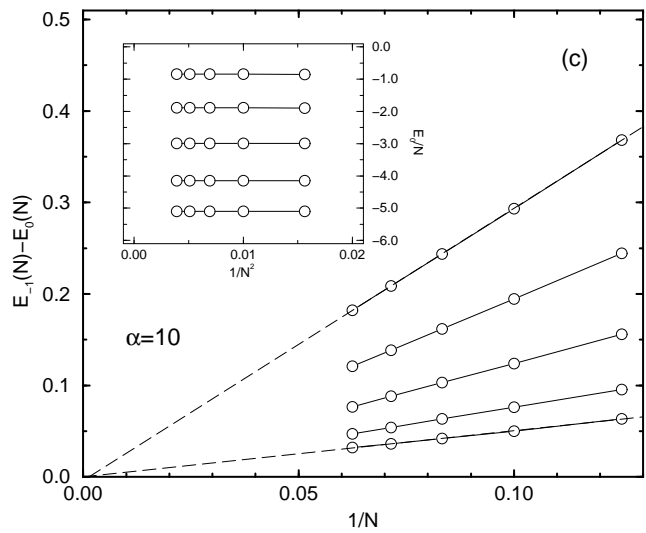
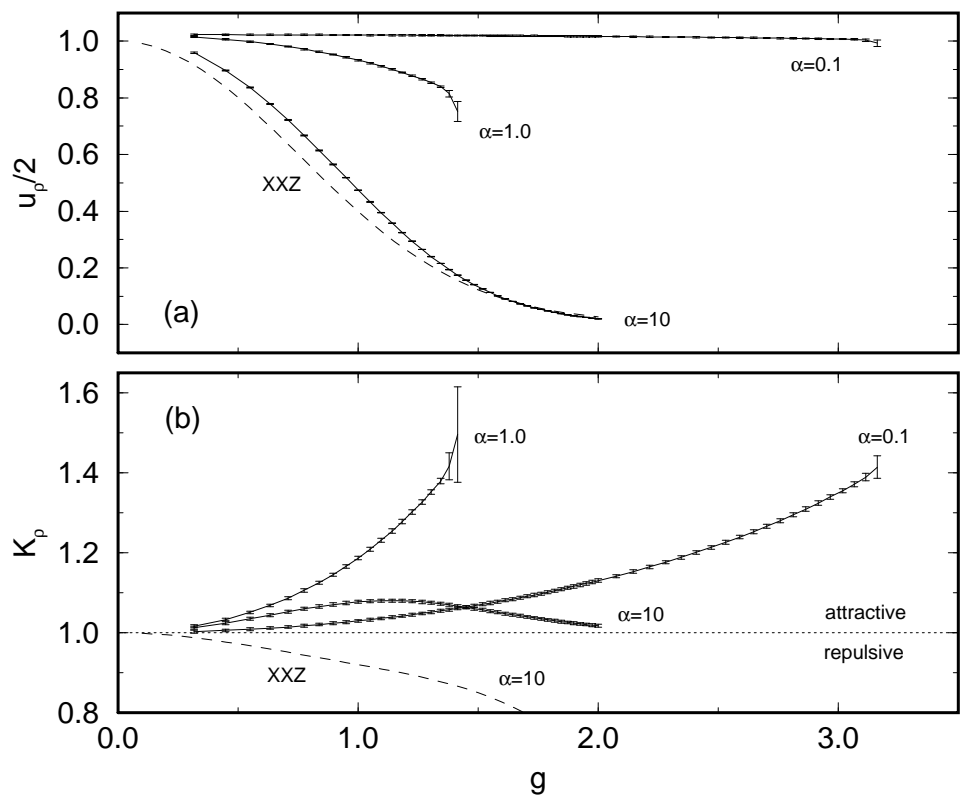
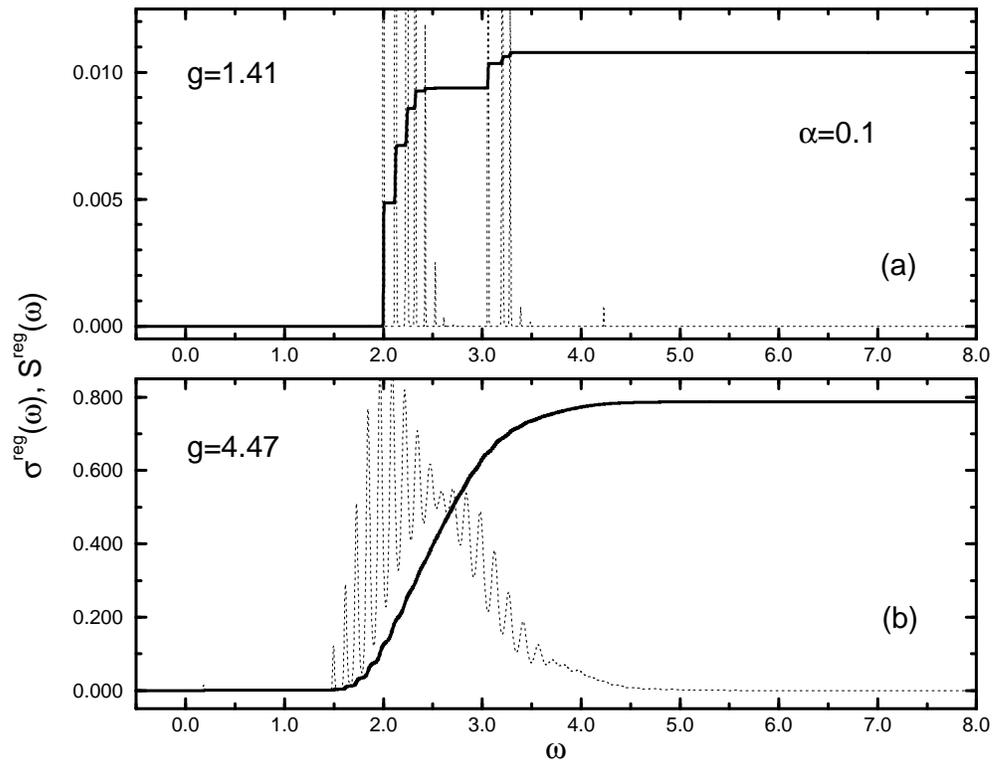
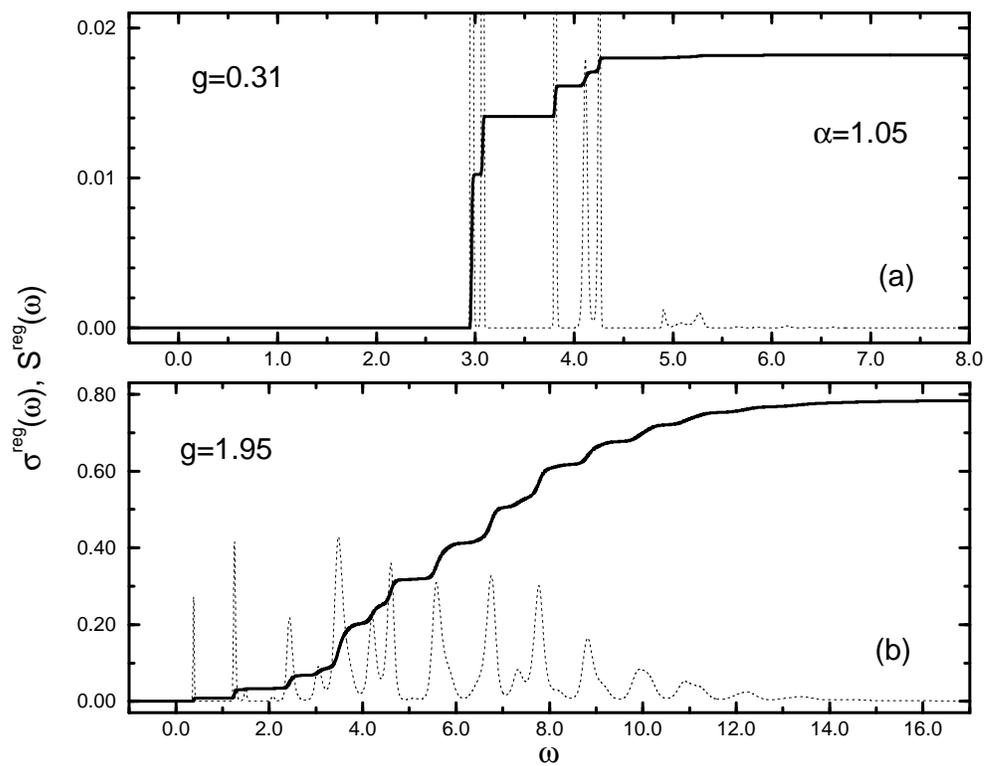
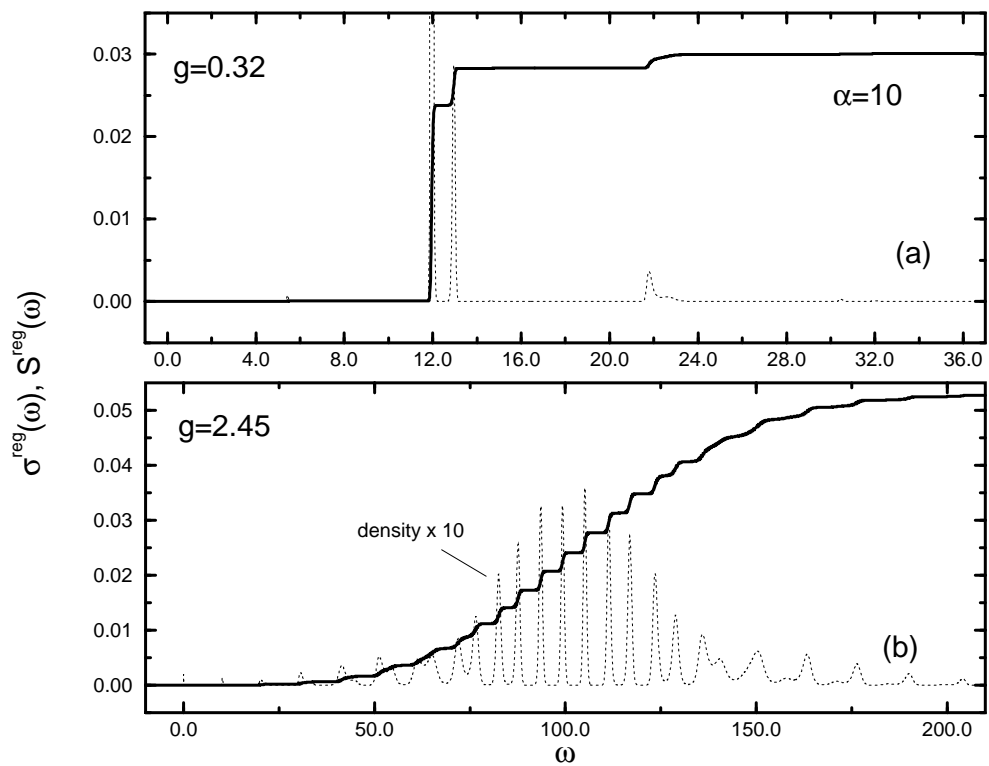
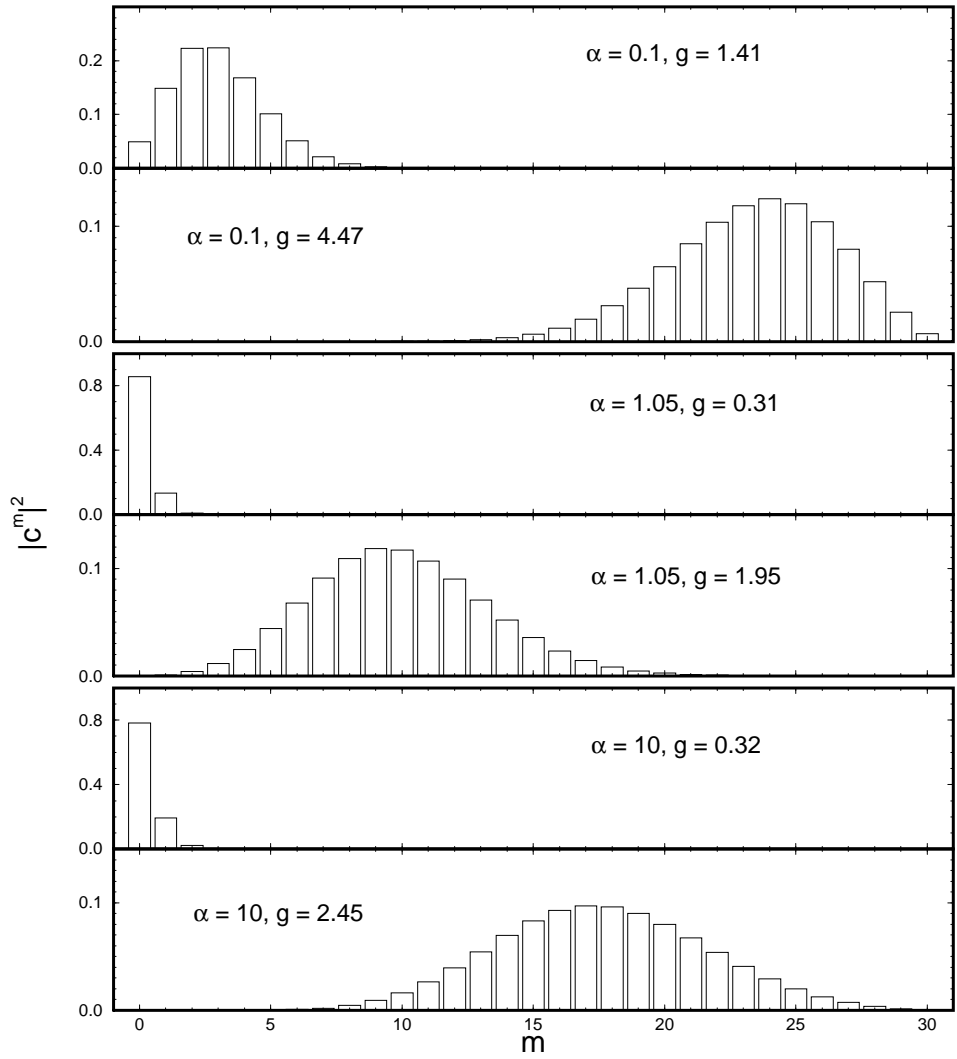
\end{document}